\documentclass[pdftex,twocolumn,epjc3]{svjour3}          
\RequirePackage[T1]{fontenc}
\smartqed  
\RequirePackage{graphicx}
\RequirePackage{mathptmx}      
\RequirePackage{flushend}
\RequirePackage[numbers,sort&compress]{natbib}
\RequirePackage[colorlinks,citecolor=blue,urlcolor=blue,linkcolor=blue]{hyperref}
\RequirePackage[utf8]{inputenc}
\RequirePackage{amsmath}

\journalname{Eur.~Phys.~J.~C}
\begin{document}

\title{Reduction of $^{222}$Rn-induced Backgrounds in a \\ Hermetic Dual-Phase Xenon Time Projection Chamber}

\author{Julia Dierle\thanksref{e1}
      \and Adam Brown 
      \and Horst Fischer 
      \and Robin Glade-Beucke 
      \and Jaron Grigat 
      \and Fabian Kuger 
      \and Sebastian Lindemann 
      \and Mariana Rajado Silva
      \and Marc Schumann\thanksref{e2}
}
\thankstext{e1}{e-mail: julia.dierle@physik.uni-freiburg.de}
\thankstext{e2}{e-mail: marc.schumann@physik.uni-freiburg.de}

\institute{Physikalisches Institut, Albert-Ludwigs-Universität Freiburg, 79104 Freiburg, Germany}

\date{Received: date / Accepted: date}

\maketitle

\begin{abstract}
The continuous emanation of $^{222}$Rn from detector surfaces causes the dominant background in current liquid xenon time projection chambers (TPCs) searching for dark matter. A significant reduction is required for the next generation of detectors which are aiming to reach the neutrino floor, such as DARWIN. $^{222}$Rn-induced back\-grounds can be reduced using a hermetic TPC, in which the sensitive target volume is mechanically separated from the rest of the detector containing the majority of Rn-emanating surfaces. We present a hermetic TPC that mainly follows the well-established design of leading xenon TPCs and has been operated successfully over a period of several weeks. By scaling up the results achieved to the DARWIN-scale, we show that the hermetic TPC concept can reduce the $^{222}$Rn concentration to the required level, even with imperfect separation of the volumes.
\end{abstract}

\section{Introduction}
\label{sec:1}

The weakly interacting massive particle (WIMP)~\cite{Feng:2010gw} is one of the most prominent candidates for the yet-unknown dark matter particle. Direct detection experiments  search for WIMPs scattering off target nuclei in low background detectors operated in deep underground laboratories~\cite{Schumann:2019eaa}. For WIMPs with masses above $\sim$3\,GeV/$c^2$ the search is currently led by dual-phase time projection chambers (TPCs) filled with cryogenic liquid xenon (LXe)~\cite{Billard:2021uyg}. Such detectors measure the prompt scintillation signal (S1) from a particle interaction, and a delayed secondary scintillation signal (S2). The latter is created by extracting the ionization electrons produced by the interaction into the gas phase above the LXe, where they generate a light signal which is proportional to the number of electrons. The time difference between S1 and S2 signal is required to drift the electrons to the liquid-gas interface and is directly related to the depth ($z$-coordinate) of the interaction.

In the WIMP search region, the background of the current generation of detectors, XENONnT~\cite{XENON:2020kmp,Aprile:2022vux}, PandaX-4T~\cite{PandaX:2018wtu,PandaX-4T:2021bab} and LZ~\cite{LZ:2019sgr,LUX-ZEPLIN:2022qhg}, is dominated by $^{214}$Pb, a progeny of $^{222}$Rn. It undergoes $\beta$-decay without emitting a coincident $\gamma$-ray with a branching ratio of $\sim$10\% leading to low-energy background events; a small fraction of these might end up in the WIMP search region. Being a daughter of the long-lived isotope $^{226}$Ra ($T_{1/2}=1602$\,y) and a noble gas, $^{222}$Rn is constantly emitted from all detector surfaces which contain traces of $^{226}$Ra. Due to its half-life of $3.82$\,d, which is long compared to the typical LXe purification time, it has sufficient time to mix with the LXe. Thus $^{214}$Bi decays can occur anywhere in the instrumented active dark matter target. Similar arguments can be made for $^{220}$Rn from the $^{232}$Th chain; however, due to its lower abundance and the much shorter half-life of 55\,s, this background source is usually suppressed compared to $^{222}$Rn.

$^{222}$Rn concentrations down to 4.5\,$\mu$Bq activity per kg of LXe have been achieved already~\cite{XENON:2020fbs}. The currently operating experiments XENONnT, LZ and PandaX-4T target concentrations of 1\,$\mu$Bq/kg. XENONnT recently published a concentration of 1.4\,$\mu$Bq/kg~\cite{Aprile:2022vux}. Future multi ton-scale detectors such as the proposed DARWIN observatory~\cite{DARWIN:2016hyl} aim for a neutrino-dominated background and thus require $^{222}$Rn concentrations of 0.1\,$\mu$Bq/kg~\cite{Schumann:2015cpa,Aalbers:2022dzr}. At this level the background from $^{222}$Rn progenies is less than 10\% of the one induced by solar pp-neutrinos in the energy region of interest. Such low concentrations shall be achieved by a combination of several methods: material selection~\cite{LZ:2020fty,XENON:2021mrg}, surface treatment (e.g., cleaning~\cite{Bruenner:2020arp} or coating) as well as online radon removal~\cite{Abe:2012,XENON100:2017gsw}. Another method is to optimize the TPC design for a low Rn~level. It is expected that only the combination of several approaches will lead to success.

One possible design approach is a hermetic TPC, where the active LXe target volume, i.e., the instrumented LXe volume inside the TPC, is mechanically separated from the LXe surrounding the TPC. While the active LXe target is only in contact with the PTFE (Teflon) reflector walls and the quartz windows of the photosensors, the outer volume is in contact with many detector components made of various materials, e.g., PTFE (TPC wall and structural elements), copper (field shaping electrodes), stainless steel/titanium (cryostat, piping, fixation), photosensor bodies, electronics components, cables, sensors etc.  A simple estimation based on the design of previous XENON detectors~\cite{XENON100:2011cza,XENON:2017lvq} suggests that the area in contact with the active target is only $\sim$10\% of the total surface area of the detector. This estimate ignores details such as cables or that the materials in contact with the outer volume usually emanate more~Rn. This implies that separating these volumes by detector construction can lead to a significant reduction of the Rn concentration in the active target volume (but only there). Ongoing independent R\&D work on this concept is reported in~\cite{Sato:2019qpr} and~\cite{Wei:2020cwl}. 

Here we present the design and operation of a kg-scale hermetic TPC prototype to demonstrate the potential of this approach. The detector features two independent gas systems to purify the inner (target) and outer LXe volumes separately. The mechanical separation between the volumes is achieved via cryofitting, i.e., the exploitation of different thermal expansion coefficients of materials. This allowed us to realize a design, presented in Sect.~\ref{sec:2}, which only differs in details from the one of successfully realized dark matter detectors. The cryogenics and Xe purification systems as well as detector operation and data taking are described in Sects.~\ref{sec:3} and~\ref{sec:4}, respectively. The hermeticity of the detector is quantified by measuring $^{83\text{m}}$Kr decays. The result for the prototype and implications for the possible $^{222}$Rn suppression at the DARWIN-scale are presented in Sect.~\ref{sec:5}.

\section{Design of the Hermetic TPC}
\label{sec:2}

\begin{figure}[ht]
    \centering
    \includegraphics[width=0.98\columnwidth]{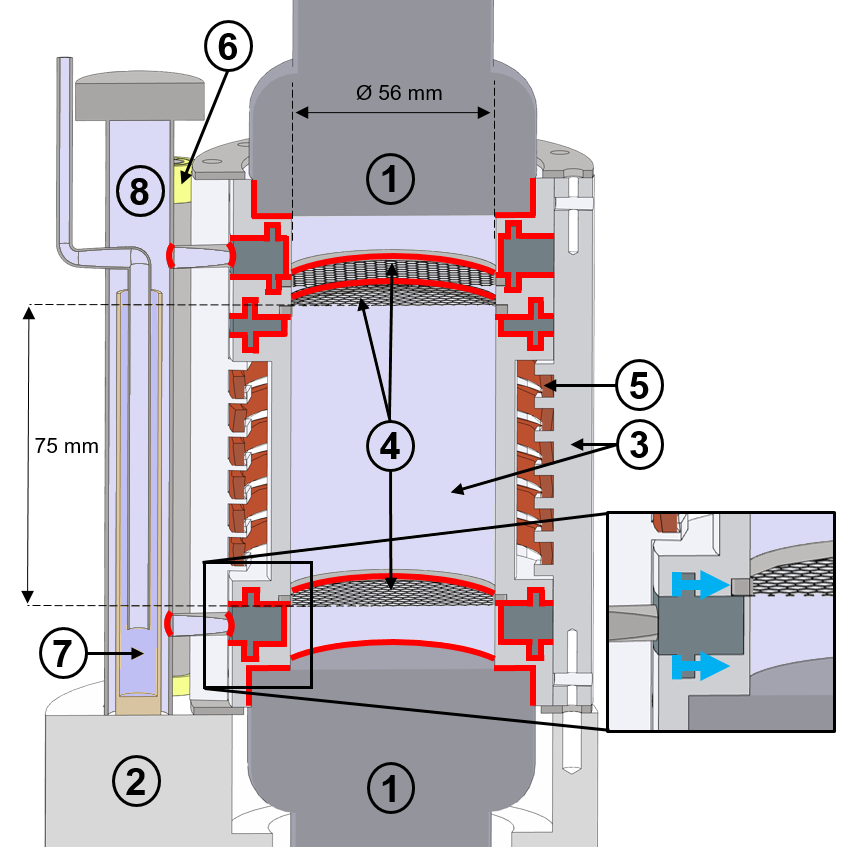} 
\caption{CAD drawing of the hermetic TPC. The inner volume is indicated by the light blue color: (1) PMTs, (2) aluminum filler, (3) PTFE reflector/structural components, (4) cathode, gate, anode electrodes, (5) field shaping electrodes, (6, partially hidden) levelmeter recording the LXe level of the outer volume, (7) weir to set the LXe level in the inner volume, installed inside a (8) stainless steel vessel. A similar vessel contains a levelmeter measuring the LXe level of the inner volume. The inset illustrates how the TPC is sealed at the interface of the PTFE chamber (white) and the cathode electrode support frame (dark gray) by means of cryofitting: when being cooled down, the shrinking PTFE exerts a force (blue arrow) onto the steel of the electrode support ring. All sealing surfaces of the TPC are marked in red. }
\label{fig:sketch}
\end{figure}

The hermetic detector prototype, shown in Fig.~\ref{fig:sketch}, is a dual-phase xenon TPC with a cylindrical active xenon target of 56\,mm diameter and 75\,mm height, enclosing $\sim$550\,g of LXe. The active region is laterally defined by a single PTFE tube of 5\,mm wall thickness which features a high reflectivity for xenon scintillation light of 175\,nm~\cite{Kravitz:2019zqv}. At the top and bottom it is closed by the quarz windows of two Hamamatsu R11410-21 PMTs of 76.2\,mm diameter. The PMTs have a quantum efficiency of 32\% at 175\,nm~\cite{Antochi:2021wik} and a collection efficiency of 95\%~\cite{Barrow:2016doe}. They are equipped with voltage divider circuits on Cirlex PCBs, which are negatively biased by a CAEN SY5527 high voltage module. The signals are transmitted via RG196 coaxial cables and digitized with a CAEN V1724 digitizer with 100\,MHz sampling rate, 40\,MHz bandwidth and 14-bit resolution.

Three mesh electrodes establish the electric drift field (between cathode and gate electrodes) and extraction field (between gate and anode) of the dual-phase TPC. The anode is installed 5\,mm above the gate electrode, which is installed 75\,mm above the cathode electrode. The meshes were etched from 0.15\,mm thick stainless steel foil and have 0.15\,mm wide webs and hexagonal 3\,mm openings, leading to $\sim$90\% optical transparency. While the gate and cathode electrodes are aligned, the anode mesh is shifted by half a mesh cell to provide more homogeneous field lines in the amplification region. The meshes are spot-welded onto stainless steel support rings. They are biased by a CAEN N1470~high voltage module. Six massive copper ring electrodes of 10\,mm height are installed at equal distances outside of the PTFE reflector tube, between the gate and cathode electrodes, and connected via 50\,M$\Omega$ \ resistors. Simulations using COMSOL Multiphysics\textsuperscript{\textregistered} v5.4~\cite{Comsol:2021upa} confirm that the electric fields in the drift and amplification regions are sufficiently homogeneous for this work.

In order to fill the sealed TPC with LXe, pipes are connected to through-holes in the anode and cathode support rings. Two more pipes in each of the frames connect to the upper and lower end of two separate cylindrical stainless steel vessels spanning the entire vertical range of the TPC, see Fig.~\ref{fig:sketch}. One of these vessels houses a capacitive levelmeter, the other one a weir. The weir's upper edge defines the LXe level inside the TPC and is fixed to 2.5\,mm above the gate level. To ease detector filling, the inner and outer LXe volumes can be connected at the height of the cathode by a custom-designed LXe valve made of PTFE. It is manually controlled via gears, rods, and a rotary motion feedthrough. To reduce the amount of xenon required to fill the detector, the body of the bottom PMT is contained in a massive aluminum filler. 

The inner target volume is separated from the outer volume by means of cryofitting, exploiting the different thermal expansion coefficients~$\alpha$ of the main TPC construction materials PTFE, stainless steel and Kovar. While PTFE has $\alpha_\mathrm{PTFE} \sim 130 \times 10^{-6}$\,K$^{-1}$, stainless steel and Kovar show a more than 10~times smaller value of $\alpha_\mathrm{SS}\sim 10 \times 10^{-6}$\,K$^{-1}$~\cite{TheEngineeringToolBox:2021tel}. Tight connections between PTFE and metal components at room temperature thus get sealed once cooled to LXe temperatures of $-100^{\circ}$C, induced by an acting force due to the larger shrinkage of PTFE (see inset of Fig.~\ref{fig:sketch}). The TPC design allows for cryofitting by embedding the three electrode frames, four pipes (both stainless steel) and the two PMT bodies (Kovar) in PTFE. While it is generally possible to achieve vacuum-tight connections with the cryofitting technique (see, e.g., \cite{Rebel:2014uia}), the limited contact area between the TPC components limited the achieved leak rates to ${\cal O}(10^{-2})$\,mbar l s$^{-1}$ in this prototype. The TPC studied here is thus not fully hermetically separated from the outside volume, however, it features only minimal modifications with respect to standard designs of LXe TPCs which are thus relatively easy to implement in a real dark matter detector.

\section{Cryogenics, Purification and Calibration}
\label{sec:3}

The hermetic TPC is operated at the versatile cryogenic detector test platform \emph{XeBRA}~\cite{Baur:2022sel}, which provides systems for xenon gas storage, cooling and purification, for data acquisition, storage and processing, as well as a slow control system for detector monitoring~\cite{Zappa:2016zsn}. The TPC is thermally decoupled from the environment by installation in a vacuum-insulated double wall cryostat equipped with multi-layer insulation.

\begin{figure}[ht]
    \centering
    \includegraphics[width=0.95\columnwidth]{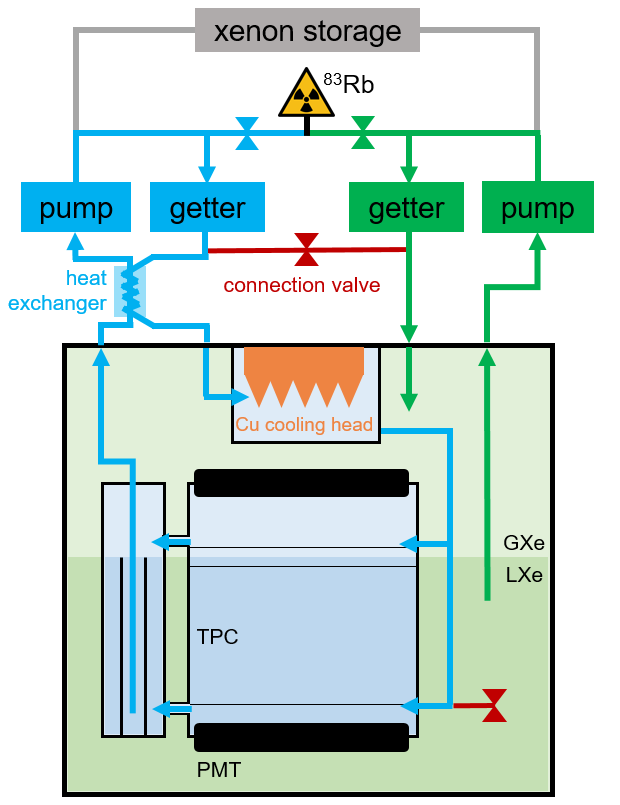}
\caption{Sketch of the xenon supply and purification systems: Xenon gas is supplied by a jointly used storage system (grey). Two independent gas systems purify the xenon contained in the inner (blue) and outer (green) volumes. The xenon is liquefied by a copper cooling head which is connected to the target volume. The volumes can be connected in their liquid and gaseous phases by valves (red). The $^{83}$Rb source installed between the two purification systems allows for detector calibration.}
\label{fig:GasSupply}
\end{figure}

The xenon gas is cooled by a copper cooling head which is in thermal contact with a liquid nitrogen reservoir. 
A cryo\-controller (Cryocon 22C) providing additional heat ensures a stable coldfinger temperature of about $-100^{\circ}$C, as required for xenon liquefaction. Temperature fluctuations are at the 0.1\,K level. Since the platform provides only one cooling system only the inner target volume is cooled actively: its gas phase is connected to the coldfinger, which is enclosed by a stainless steel cap to separate it from the surrounding gas of the outer volume (see Fig.~\ref{fig:GasSupply}). The xenon in the outer volume is cooled only indirectly via thermal contact. The LXe valve is open during detector cool-down/filling to ensure a fast thermalization time.

The mechanical separation of the two volumes requires two  gas systems for independent xenon gas supply and purification. Both systems are made of 1/4-inch stainless steel pipes connected by Swagelok VCR components. A sketch of the gas routing is shown in Fig.~\ref{fig:GasSupply}. The two systems are connected to the same xenon storage unit which supplies the gas for the initial filling. During detector operation, when the LXe valve is closed, the two gas systems independently purify the xenon gas from the two volumes in hot getters (SAES MonoTorr PS3-MT3-R-2); the purification flows are driven by membrane pumps (KNF PM 28544-022). To purify the target volume, LXe is extracted from the weir via a heat exchanger. The purified gaseous xenon (GXe) is returned via the heat exchanger, which partly liquefies the GXe, into the closed volume around the cooling head. The liquefied xenon formed here is collected by a funnel and guided into the target volume at the cathode level. To purify the outer volume, xenon is extracted from the liquid phase and the purified GXe is returned to the gas phase. The gaseous phases of the separated xenon volumes can be connected by a valve installed outside the cryostat; this valve is closed during operation in hermetic mode. 

The detector is calibrated by injecting atoms of a $^{83\text{m}}$Kr conversion electron source into the LXe. It provides a delayed coincidence signature of two low-energy signals at 32.1\,keV and 9.4\,keV~\cite{Kastens:2009pa,Manalaysay:2009yq}. The mother isotope $^{83}$Rb is contained in zeolite beads, which are installed behind by a micrometer~filter~\cite{Hannen:2011mr} in a VCR tee. This is installed between the two purification systems such that $^{83\text{m}}$Kr can be injected into one (or both) of the xenon volumes. Being a noble gas, $^{83\text{m}}$Kr is not removed by the getters. However, its short half-life of 1.83\,h ensures that it decays quickly after closing off the $^{83}$Rb source.

\section{Detector Operation and Data Taking}
\label{sec:4}

After the initial filling of the detector with LXe, the LXe control valve connecting inner and outer volume was closed. Stable long-term detector conditions in this "hermetic mode" could be achieved when only purifying the LXe in the inner target volume. Additional purification of the outer volume, even at very low flows, led to liquid-level fluctuations in the inner volume at the mm-scale, which prevented proper TPC operation since it severely affected the signal quality. This behavior is attributed to the absence of a dedicated cooling head for the outer volume: the cooling of the gas in the outer volume via thermal contact to the inner one is not sufficient to compensate for the extra heat load introduced by the purification system. It seems that the operation of a hermetic detector with independent purification of both volumes requires independent cooling heads with temperature control systems for both volumes or at least efficient heat exchangers in both purification system. For the purpose of this work, however, the purification of the LXe in the outer volume is not required. The purification of the inner target volume works well, reaching electron lifetimes of up to 300\,$\mu$s, significantly exceeding the maximum electron drift time of 43\,$\mu$s.

The data presented here were acquired with the cathode biased at $-$4\,kV and the gate at ground, establishing an average drift field of 530\,V/cm. The anode was biased at $+$4\,kV. The data were recorded using a triggerless readout framework adapted from XENON~\cite{XENON:2019bth} and stored on a server. The data is processed in the following way: after the initial identification of peaks as excursions from the baseline, S1- and S2-like signals are identified based on their area and width, where the width is defined as the time interval containing 50\% of the central peak area. Signals are merged into events if they occur within a pre-defined time window, given by the maximal electron drift time across the TPC. 

The gains required to convert the PMT signals into photoelectrons (PEs) are measured at the beginning of each data taking campaign by using an externally trigged blue LED ($\sim$460\,nm), fixed to the TPC structure outside the target volume. A pulse generator (RIGOL DG1022) is used to control light intensity and pulse rate. The PMT gains were stable within 4\% during extended measurement periods. 

\begin{figure}[ht]
    \centering
    \includegraphics[width=\columnwidth]{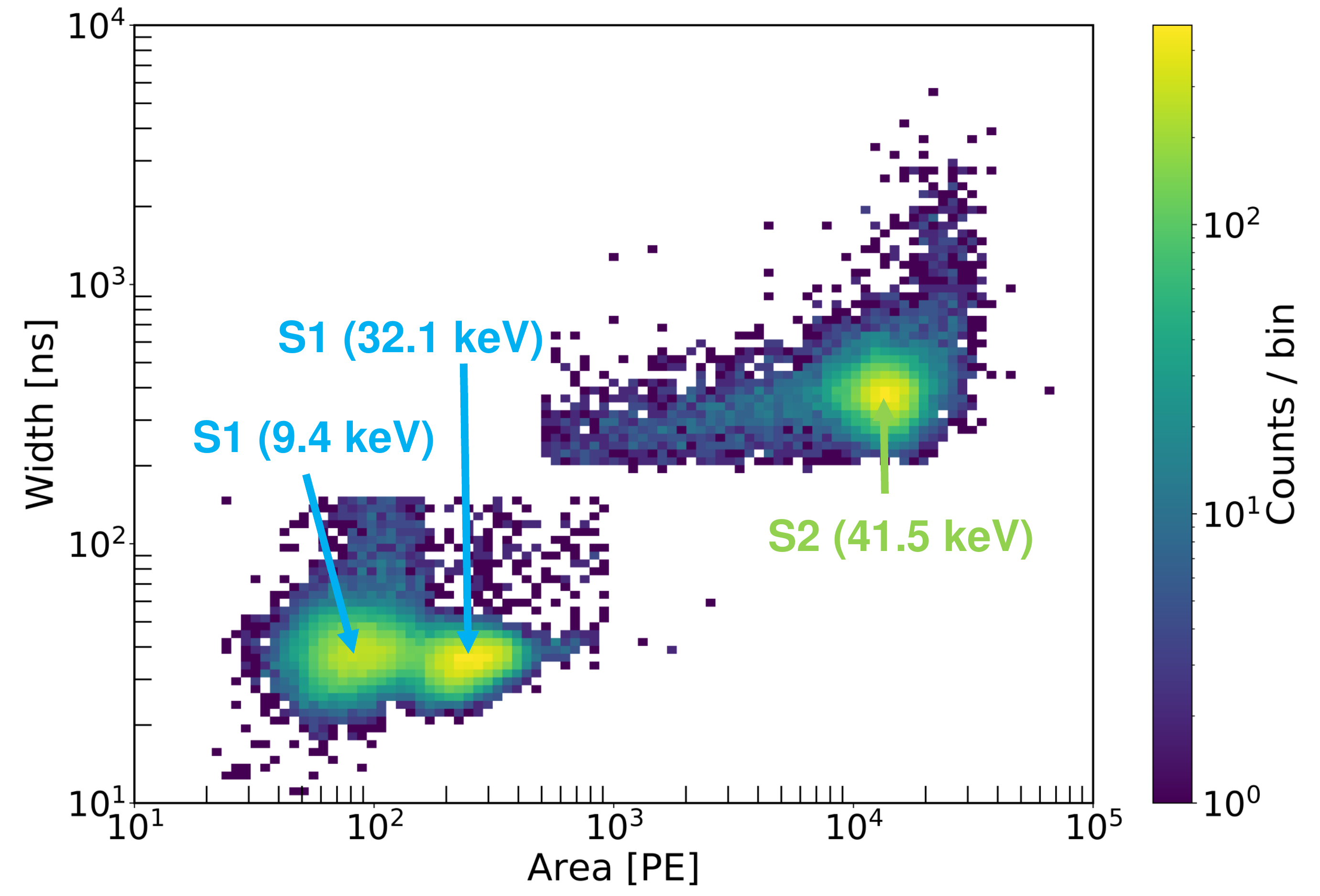}
\caption{Histogram of the individual peaks contained in events selected as originating from $^{83\text{m}}$Kr: the two populations at smaller signal width correspond to the 9.4\,keV and 32.1\,keV S1~signals, respectively. The S2 signals from the summation peak at 41.5\,keV are visible at larger area and width. }
\label{fig:s1s2}
\end{figure}

The evaluation of the TPC hermeticity presented in Sect.~\ref{sec:5} relies on the injection of $^{83\text{m}}$Kr into the detector, rendering the identification of $^{83\text{m}}$Kr events in the data of central importance. Exploiting the delayed coincidence signature of its decays, the following selection criteria are used to select a clean $^{83\text{m}}$Kr data set: (i) A valid event must contain at least three peaks with at least two~S1 and one S2~peak. Only one S2 is required since the wider S2~signals of the two decays usually overlap. (ii) The largest S1~peak must occur before the second-largest S1~peak since the 32.1\,keV decay happens first. (iii) The area of both S1 peaks must be at least 20\,PE and the S2 area at least 500\,PE. (iv) The event's $z$-position, defined by the time difference between the largest S1 and the largest S2 signal, must fall in the central part of the TPC, more than 10\,mm away from the gate and cathode electrodes. The individual peaks constituting the selected $^{83\text{m}}$Kr events are shown in Fig.~\ref{fig:s1s2}: the two populations at lower peak area and width correspond to the S1~peaks; the population at larger area and width is from the overlapping S2 signals with a total energy of 41.5\,keV. The validity of the $^{83\text{m}}$Kr selection was verified by measuring the half-life of the short-lived intermediate state: with $T_{1/2} = (158 \pm 4)\,$ns it agrees very well with the literature value of 156.8\,ns~\cite{McCutchan:2015vcl}. 

The S1 signal areas are corrected for the $z$-dependence of the light collection efficiency which is mainly caused by the varying solid angle coverage and total reflection on the liquid-gas interface. The correction, a second-order polynomial, is derived using $^{83\text{m}}$Kr events. The detector-specific signal yields $g1 = (0.089 \pm 0.004)$\,PE/$\gamma$ and $g2 = (7.1 \pm 0.7)$\,PE/$e^-$ are obtained by measuring the $^{83\text{m}}$Kr light and charge signals at different drift fields. The yields are comparable to other small- and large-scale TPCs, e.g., ~\cite{XENON:2017sic, LUX:2015amk, Hogenbirk:2018knr}, demonstrating the good performance of the hermetic TPC prototype. However, we note that neither these yields nor the detector bias voltages impact the hermeticity study as long as $^{83\text{m}}$Kr events can be unambiguously identified.

\section{Results and Discussion}
\label{sec:5}

In this section we evaluate the hermeticity of the fully functional prototype detector and scale up the result to estimate the impact of a hermetic TPC on the $^{222}$Rn background of a future multi ton-scale LXe-based dark matter detector such as DARWIN.

\subsection{Hermeticity of the Prototype TPC}
\label{sec:5_1}

To determine the level of hermeticity of the prototype TPC, $^{83\text{m}}$Kr was only injected into the inner LXe target volume and the rate of $^{83\text{m}}$Kr decays was recorded over time. The decreasing trend from the expected exponential decay will be enhanced by an additional mass exchange flow~$f$ (defined as mass per unit time) transporting $^{83\text{m}}$Kr atoms dissolved in the xenon to the outer volume, where they will decay unobserved. 

\begin{figure}[ht]
    \centering
    \includegraphics[width=\columnwidth]{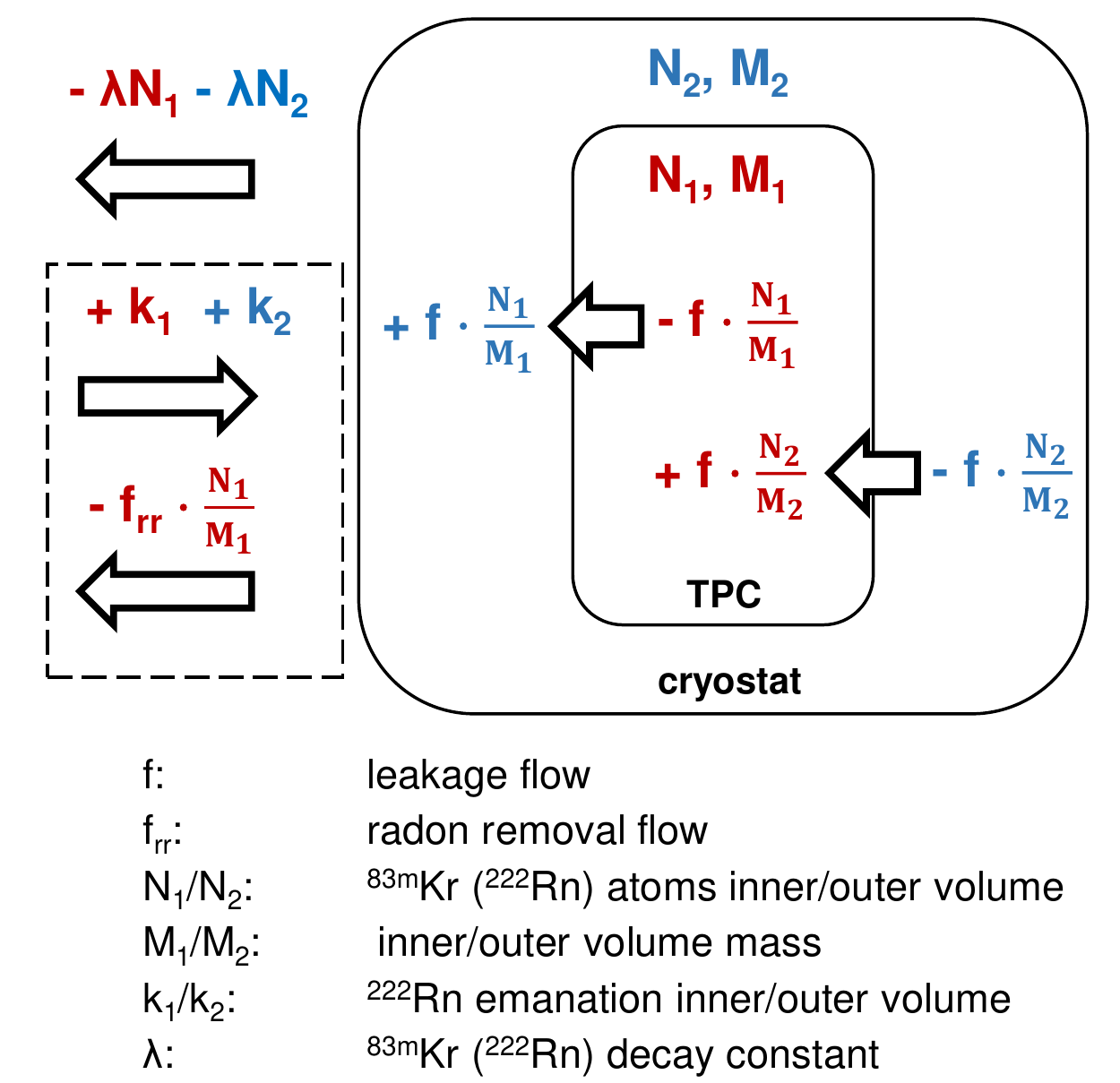} 
\caption{Sketch of the various terms to describe the evolution of an impurity concentration (here: $^{83\text{m}}$Kr or $^{222}$Rn) in an hermetic TPC. The red (blue) terms represent factors influencing the evolution of the number of impurity atoms~$N_1$ ($N_2$) in the inner (outer volume). The terms in the dashed box describe the constant emanation of impurities into the system and their removal in a purification system and only apply for the discussion of the $^{222}$Rn concentration in DARWIN.}
\label{fig:KryptonRadonModel}
\end{figure}

The mass flow~$f$ is derived using a model that describes the components affecting the $^{83\text{m}}$Kr concentration in both xenon volumes, illustrated in Fig.~\ref{fig:KryptonRadonModel}: $N_1$ ($N_2$) and $M_1$ ($M_2$) denote the number of krypton atoms and the xenon mass of the inner (outer) volume, respectively. 
A homogeneous krypton concentration is assumed within each of the volumes. The xenon purification system is neglected due to the low xenon mass and consequently low number of krypton atoms in the pipes of the gas purification system. Since the total amount of xenon in the system remained constant, and since the system showed very good long-term stability with only minor pressure differences between the two xenon volumes, the absolute leakage flow in both directions is assumed to be equal, i.e., $|f| = |f_{out}| = |f_{in}|$. 

Three terms describe the sources and sinks that affect the krypton concentration in the LXe target after the initial $^{83\text{m}}$Kr injection: \newline
(i) $- \lambda N_1$: The number of $^{83\text{m}}$Kr atoms contained in the target volume decays with the decay constant~$\lambda$. \\
(ii)$- f \frac{N_1}{M_{1}}$: The number of $^{83\text{m}}$Kr atoms contained in the target volume is decreased by the leakage flow $f$ from the inner to the outer volume, scaled by the krypton concentration $N_1/M_1$ in the inner volume. \\
(iii) $+ f \frac{N_2}{M_{2}}$: The number of $^{83\text{m}}$Kr atoms contained in the target volume is increased by the leakage flow $f$ from the outer to the inner volume, scaled by the krypton concentration $N_2/M_2$ in the outer volume. 

Similar terms describe the krypton concentration in the outer volume (see Fig.~\ref{fig:KryptonRadonModel}). The model can be easily expanded to include effects such as a target purification or constant replenishment of the source, factors that are relevant to describe $^{222}$Rn in DARWIN, see Sect.~\ref{sec:5_2} below. The full system is described by two coupled differential equations: \begin{align}
\frac{\partial N_1}{\partial t} &= -\lambda N_1 - f  \frac{N_1}{M_{1}} + f  \frac{N_2}{M_{2}} \\
\frac{\partial N_2}{\partial t} &= -\lambda N_2 - f  \frac{N_2}{M_{2}} + f  \frac{N_1}{M_{1}} \:\: .
\end{align}

\begin{figure}[t]
    \centering
    \includegraphics[width=\columnwidth]{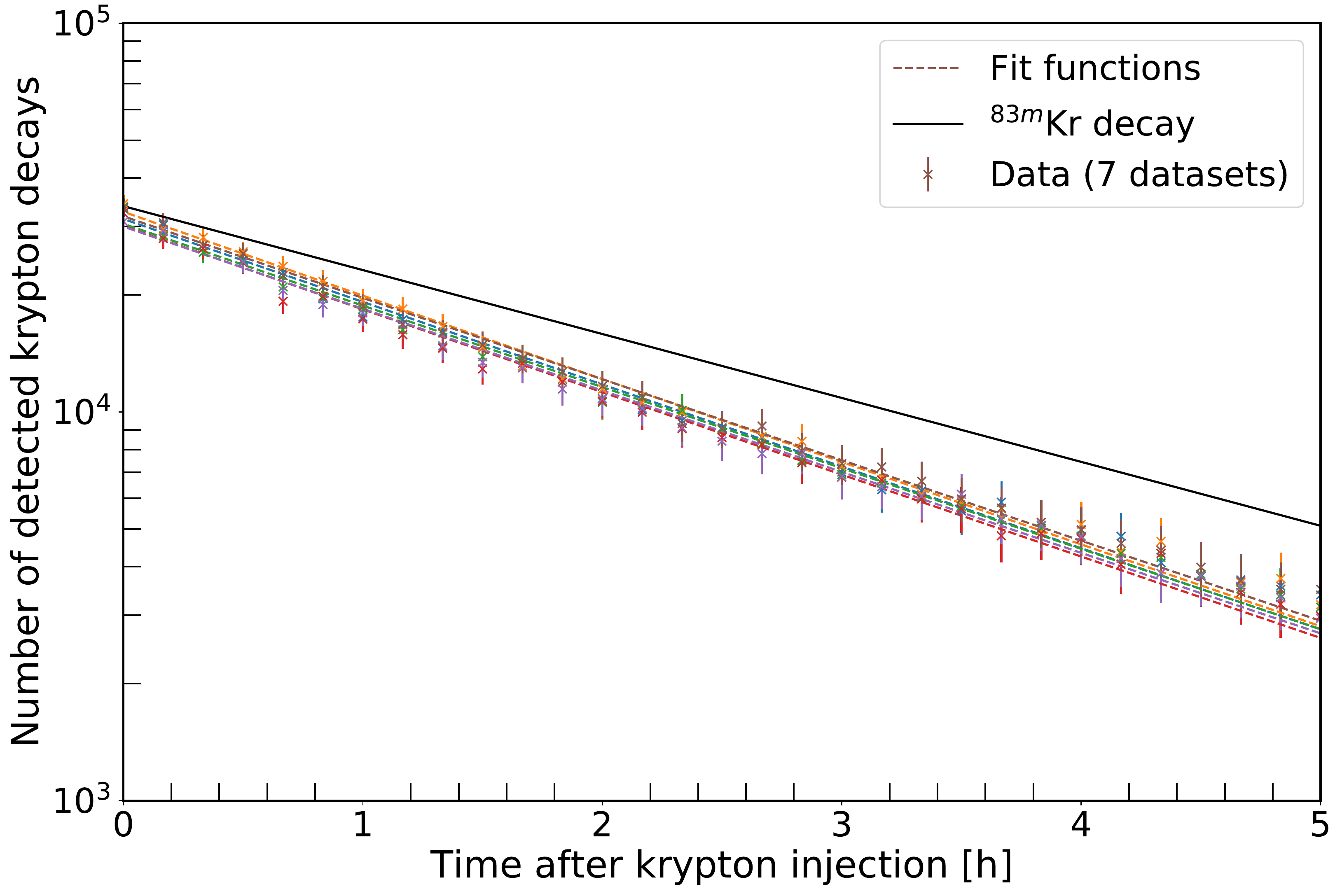}
\caption{Number of measured decays per 3~min interval vs.~time after injecting $^{83\text{m}}$Kr into the inner volume. Seven individual measurements are shown. The black curve describes the expected exponential decrease for a fully sealed detector. The observed deviation is caused by a finite leakage flow~$f$ which additionally reduces the number of $^{83\text{m}}$Kr atoms. $f$~can be extracted from a fit of the solution $N_1(t)$ of the differential equations~(1) and~(2) to the data. }
\label{fig:KrResults}
\end{figure}

The solution $N_1(t)$ to these equations depends on the leakage flow~$f$. The initial conditions are defined by the $^{83\text{m}}$Kr injection procedure: $N_1(t = 0) = N^i$ and $N_2 (t = 0) = 0$, where $N^i$ denotes the unknown number of $^{83\text{m}}$Kr atoms initially filled. Since the number of krypton decays, the activity $A_1(t)$, is directly proportional to $N_1(t)$, this function can be used to describe the measured data.
Fig.~\ref{fig:KrResults} shows seven independent measurements of  detected $^{83\text{m}}$Kr decays in 3\,min intervals after the krypton injection. All datasets were individually described by $N_1(t)$ to extract the leakage flow~$f$; the mean value for the prototype is $f_\textrm{p} = (0.11 \pm 0.01)$\,kg/h. For comparison, Fig.~\ref{fig:KrResults} also shows the expected behavior for a fully hermetic TPC, where the number of krypton atoms contained in the inner volume is only affected by the radioactive decay. The measured flow $f_\textrm{p}$ is several times lower than typical purification flows in TPCs of standard design, where the xenon is usually extracted from the outer volume and returned (after purification) into the inner target, i.e., the purification flow essentially acts as the leakage flow~$f$.

This promising result shows that the exchange of LXe and thus the amount of radioactive impurities can be reduced by a hermetic TPC design exploiting the cryofitting technique, with minimal deviations from existing detector design concepts. Improvements could be achieved, e.g., by optimizing the sealing contact surfaces or by sealing the top and bottom of the TPC with large quartz plates while operating the PMTs in the outer volume (see, e.g.,~\cite{Sato:2019qpr}).

\subsection{Possible $^{222}$Rn reduction in DARWIN}
\label{sec:5_2}

Due to the very different TPC sizes, the hermeticity of the prototype TPC presented above needs to be scaled up to make predictions of the $^{222}$Rn reduction in DARWIN with this approach. The scaling factor depends on the origin of the main leakage. If this origin were around the PMTs, the measured leakage flow~$f_p$ would have to be scaled with the TPC top/bottom area $\propto r^2$, or a factor~1000. A leak around the electrode frames would scale with the radius~$r$ or a factor of~50. A leak around the LXe valve or the liquid level control would not require any significant scaling. Based on the data acquired in this work it is not possible to distinguish between these cases. Thus, the estimations of the reduction of the $^{222}$Rn concentration in DARWIN using a hermetic TPC design are presented below as a function of the leakage flow~$f$, highlighting these three scenarios. It is also possible that the observed leak is at the vessel surrounding the coldfinger, required to push the purified and liquefied gas in the inner volume. This case could be mitigated by straightforward design modifications which do not affect the TPC itself and is thus not considered in the following. 

The impurity model introduced in equations~(1) and~(2) to describe the $^{83\text{m}}$Kr evolution in the prototype detector needs to be adapted to treat $^{222}$Rn. This is constantly emanated from all detector surfaces but can also be actively removed by dedicated Rn-removal systems~\cite{XENON100:2017gsw,Cui:2020bwf}, see Fig.~\ref{fig:KryptonRadonModel}. The advantage of a hermetic TPC is that the purification efforts can concentrate on the smaller and cleaner LXe target. Due to the $3.8$\,day half-life of $^{222}$Rn~\cite{Bellotti:2015pmo}, homogeneous mixing of radon within the LXe target is expected even for a large detector such as DARWIN.  $M_1$, $N_1$ and $M_2$, $N_2$ denote again the xenon mass and number of radon atoms in the inner (1) and outer volume (2), respectively. The two additional terms influencing the radon concentration in the target volume are: \newline
(iv) $+ k_i$: Radon is constantly emanated from all surfaces in the detector which is described by the emanation rates $k_1$ and $k_2$ for the inner and outer volume, respectively. \\
(v) $- f_{rr} \frac{N_1}{M_{1}}$: The radon removal system (for the inner target) with a purification flow $f_{rr}$ is assumed to have a 100\% removal efficiency. No radon removal system is assumed for the outer volume.

Combining all terms to describe the full system yields:
\begin{align}
\frac{\partial N_1}{\partial t} &= -\lambda N_1 - f  \frac{N_1}{M_{1}} + f  \frac{N_2}{M_{2}} + k_1 - f_{rr}  \frac{N_1}{M_1}\\
\frac{\partial N_2}{\partial t} &= -\lambda N_2 - f  \frac{N_2}{M_{2}} + f  \frac{N_1}{M_{1}} + k_2 \:\: .
\end{align}
The analytic solution $N_1(t)$ to this set of differential equations is evaluated at $t \to \infty$, where the radon concentration reaches its equilibrium. The radon activity $A_1$ in the target volume is obtained by multiplying $N_1(t \to \infty)$ with the decay constant $\lambda$. 

\begin{figure}[t!]
    \centering
    \includegraphics[width=\columnwidth]{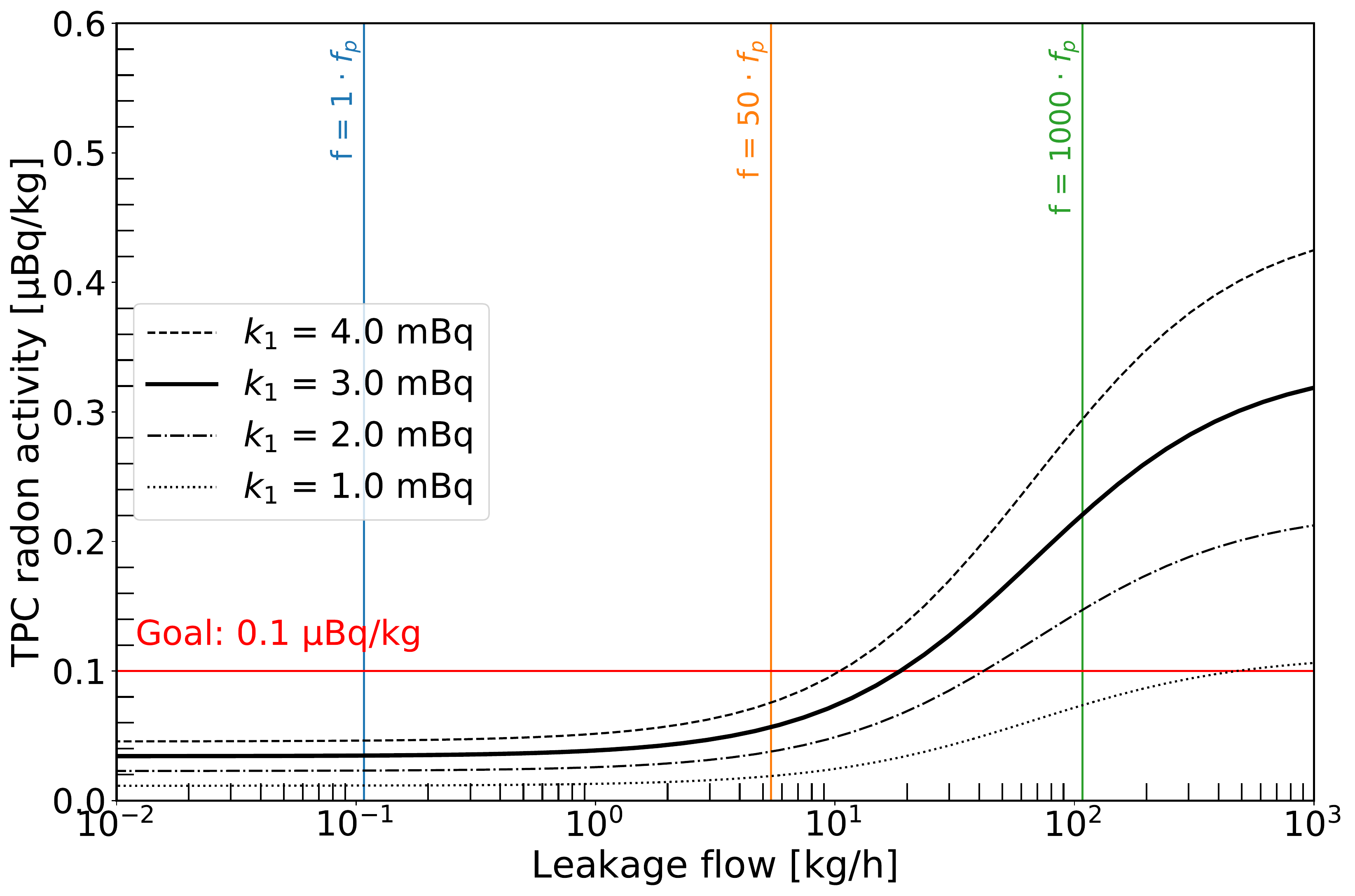} \\
    \includegraphics[width=\columnwidth]{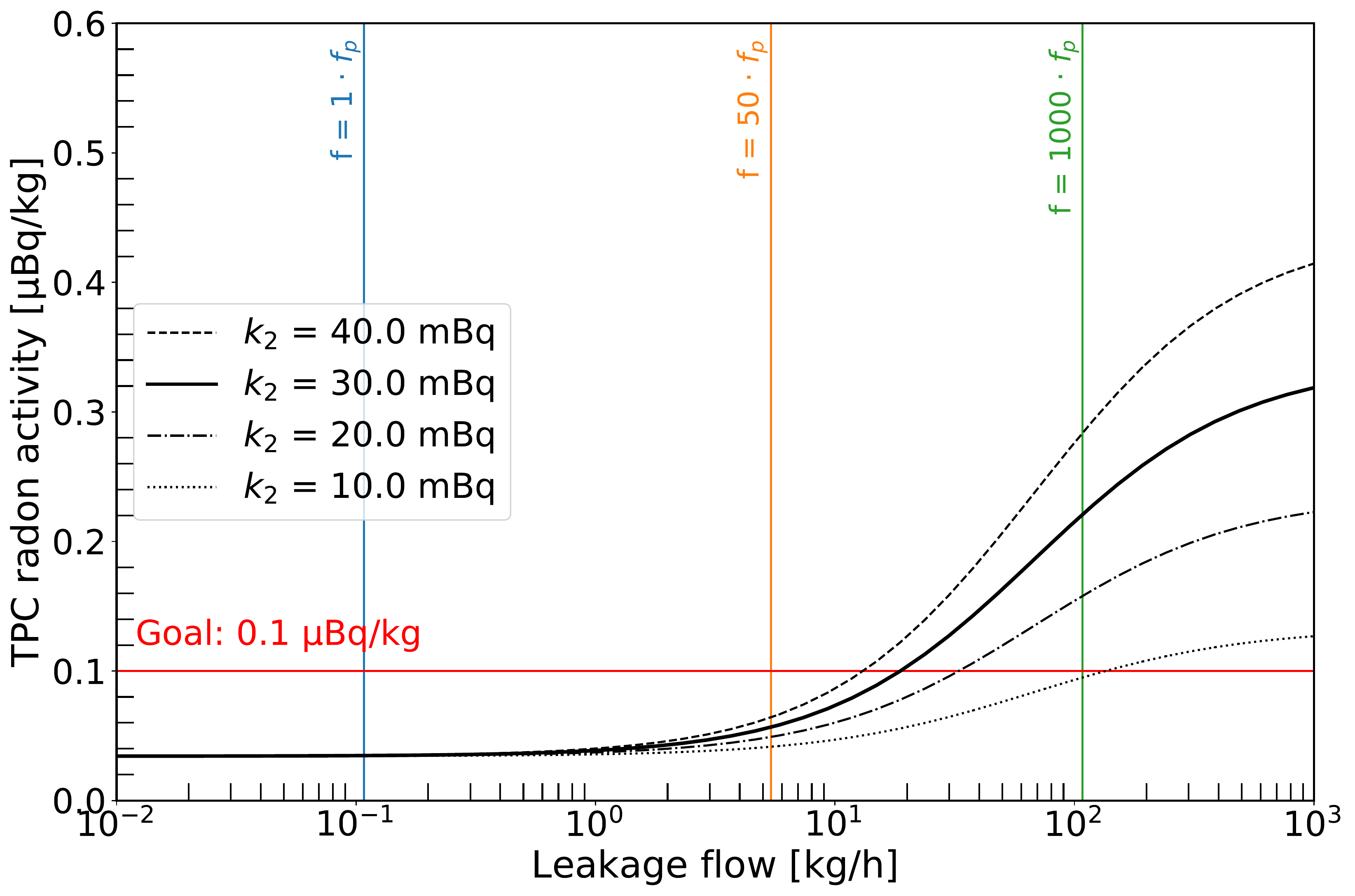} \\
    \includegraphics[width=\columnwidth]{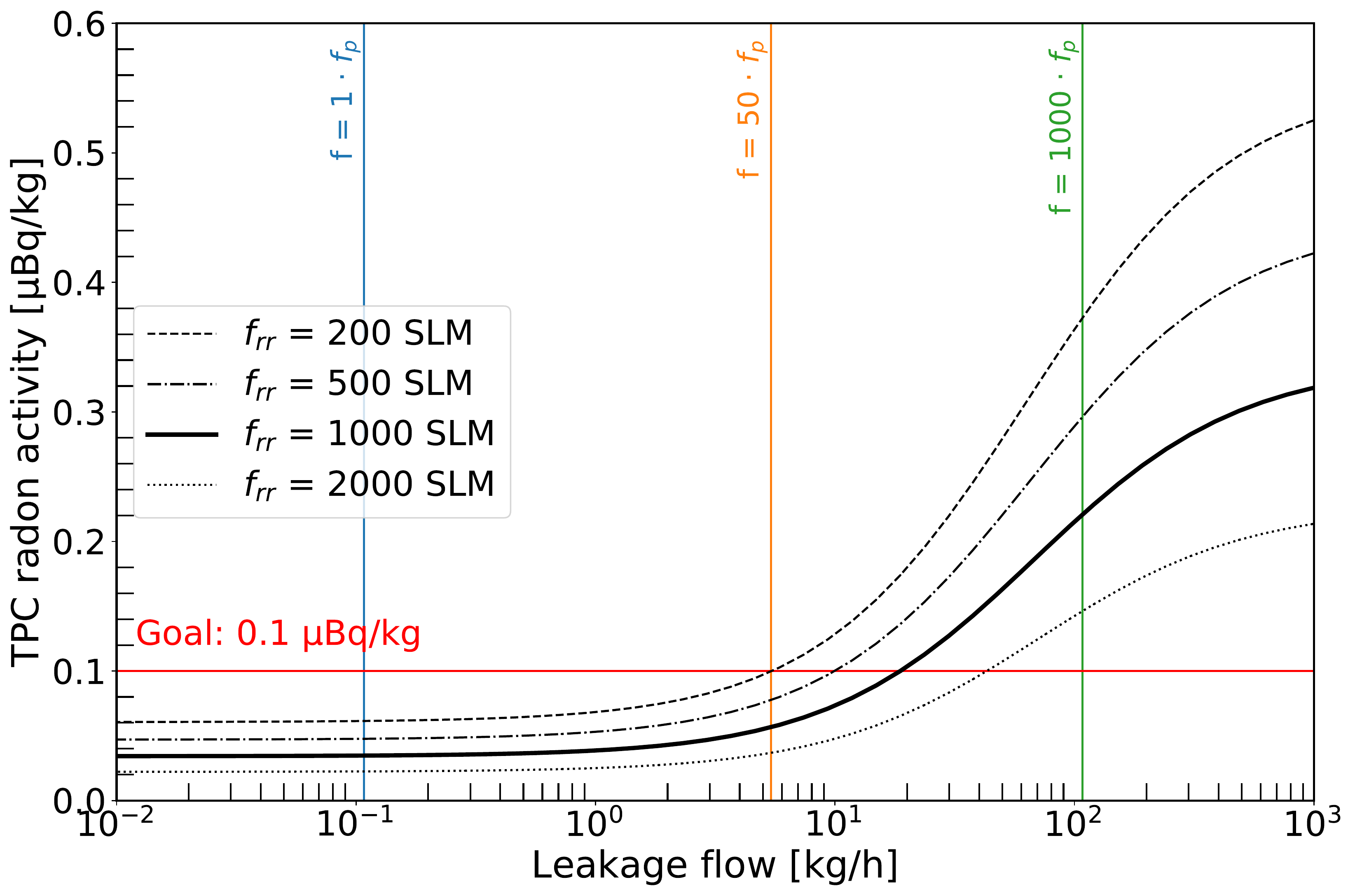} 
\caption{Impact of detector parameters on the expected specific $^{222}$Rn activity in DARWIN, for different levels of hermeticity defined by the leakage flow. The thick black line is identical in all panels and corresponds to $^{222}$Rn emanation rates $k_1=3$\,mBq and $k_2=30$\,mBq in the inner and outer volumes, respectively, and a feeding flow $f_{rr}=1000$\,SLM to a radon removal system. One of these parameters is varied in the individual panels (top: $k_1$, center: $k_2$, bottom: $f_{rr}$) while the other two are kept constant. Three different ways of scaling up the hermeticity result of the prototype detector (see text) are indicated by the vertical lines. All scenarios show the great potential of a hermetic TPC to achieve the $^{222}$Rn concentration goal of 0.1\,$\mu$Bq/kg in DARWIN (red line), especially in combination with other radon reduction efforts.}
\label{fig:RadonModelDARWIN_ChangingEmanation}
\end{figure}

We assume the following baseline parameters to study the impact of a hermetic detector (with a given leakage flux~$f$) on the $^{222}$Rn budget: total radon emanation rates of $k_1 = 3$\,mBq and $k_2 = 30$\,mBq in the inner and outer volume, respectively, and a xenon flow $f_{rr} = 1000\,\text{SLM} \approx 360\,\text{kg/h}$ feeding the radon removal system. These numbers are based on radon emanation measurements from XENON1T~\cite{XENON:2020fbs}, the DARWIN benchmark design with $M_1 = 40\,$t and $M_2 = 10$\,t~\cite{Schumann:2015cpa,DARWIN:2016hyl}, and assuming 10~times more surface area in the outer volume (estimated based on the design of several XENON detectors). The resulting specific $^{222}$Rn activity depends on the level of hermeticity described by~$f$ and is shown in Fig.~\ref{fig:RadonModelDARWIN_ChangingEmanation} (solid line on all panels): High leakage flows, representing the case of a classical, non-hermetic TPC, yield a radon activity which is more than three times higher than the DARWIN goal of 0.1\,$\mu$Bq/kg~\cite{Schumann:2015cpa,DARWIN:2016hyl,Aalbers:2022dzr}. In contrast, the low leakage flows of a very well sealed TPC lead to $^{222}$Rn concentrations of less than $0.05\,\mu$Bq/kg. This showcases the great potential for radon reduction offered by the hermetic TPC concept. Note that even a semi-hermetic TPC would significantly impact the achievable radon background level: a TPC with a leakage flow of $f = 50 f_\textrm{p}$ reduces the $^{222}$Rn concentration by a factor $\sim$6 compared to a non-sealed design and still outperforms the radon goal. Obtaining such a leakage flow is not unrealistic, as demonstrated with the prototype and in particular after further optimizing the sealing areas. 

The results discussed so far depend on assumptions regarding parameters which describe the performance of other radon mitigation techniques: the emanation rates~$k_1$ and $k_2$ represent the impact of  material selection and surface treatment, while the flow $f_{rr}$ quantifies the impact of radon distillation systems. Their influence is studied by individually varying one of these parameters $k_1$, $k_2$, $f_{rr}$ while fixing the others to $k_1=3$\,mBq, $k_2=30$\,mBq, or $f_{rr}=1000$\,SLM. Fig.~\ref{fig:RadonModelDARWIN_ChangingEmanation}\,(top) shows the specific radon activity for four different radon emanation rates $k_1$ in the inner volume. Each scenario reaches $^{222}$Rn concentrations below the DARWIN design goal for a hermetically tight TPC, while a fully open TPC design ($f\to \infty$) results in excessive $^{222}$Rn activity in most cases. The impact of different $^{222}$Rn emanation rates $k_2$ in the outer volume is shown in Fig.~\ref{fig:RadonModelDARWIN_ChangingEmanation}~(center): even the most optimistic assumption, $k_2 = 10\,$mBq, leads to radon activities above the DARWIN design goal for classical non-hermetic TPCs. 
Fig.~\ref{fig:RadonModelDARWIN_ChangingEmanation}~(bottom) presents the impact of the  flow~$f_{rr}$ feeding a radon removal system: $f_{rr}=2000$\,SLM would not be sufficient to reach DARWIN’s design goal in a classical TPC design. A flow $f_{rr} \approx 6000$\,SLM would be required, corresponding to purifying the entire xenon inventory once per day. The power consumption of a distillation-based radon removal system scales linearly with its feeding flow $f_{rr}$ and thus results in a very cost-intensive radon mitigation technique. Again, a semi-hermetic detector with $f = 50 f_\textrm{p}$ would allow reaching the DARWIN design goal with relatively modest feeding flows.

\section{Summary}
\label{sec:6}

The reduction of $^{222}$Rn backgrounds to a level well below the contribution from low-energy solar neutrinos is essential, but challenging for next-generation LXe-based WIMP dark matter detectors such as DARWIN. A hermetic TPC separates the inner LXe target from the outer LXe volume surrounding the detector and thereby reduces the surface area contributing to the radon background. This promising concept is one of a few possibilities to reduce radon backgrounds. Most likely, a combination of several methods will be required to reach the $^{222}$Rn background goal. 

A small-scale hermetic TPC prototype was built and operated to demonstrate the potential of this concept. It features two gas systems for the independent purification of the two xenon volumes. However, the lack of an independent cooling system for the outer volume prevented the operation of the hermetic TPC with both volumes being purified. Data presented here were acquired with only purifying the inner target volume. The hermetic TPC was operated stably over weeks, with light and charge yields comparable to other detectors. Electron lifetimes exceeding the the maximum drift length by almost an order of magnitude were achieved.

The TPC design is based on the well-established, successful dual-phase LXe TPCs for dark matter searches. It was optimized for the hermetic operation by mechanically mating different components defining the boundary between the two LXe volumes using the principle of cryofitting, exploiting different thermal expansion coefficients of PTFE and metals. The level of hermeticity can be quantified by the leakage flow~$f$ between two volumes. In the prototype it was measured as $f = (0.11 \pm 0.01)$\,kg/h by monitoring the decrease in activity of homogeneously dissolved $^{83\text{m}}$Kr isotopes in the inner volume. This flow is several times smaller than for classical TPCs of similar size and demonstrates that (semi)-hermetic TPCs can be successfully operated.

The expected $^{222}$Rn concentration in a DARWIN-scale detector optimized for hermetic operation depends on various parameters, which describe the $^{222}$Rn contamination in the two volumes, and was calculated for different leakage flows~$f$. The result from the hermetic prototype TPC was scaled up to the DARWIN dimensions assuming three different scenarios for the unknown origin of the leakage. A hermetic TPC can reduce the $^{222}$Rn concentration to the level required for a neutrino-back\-ground-dominated dark matter detector, even if no full hermeticity can be achieved. 

The operation of two independent gas systems for both volumes and the mechanical separation between them is a relatively easy method and can complement other radon reduction methods. Future studies need to explore whether the very (radio-)clean cryofitting technique employed here can be reliably used for larger TPCs or whether alternative sealing methods need to be investigated. The level of hermeticity of current large-scales detectors is not known but their leakage flow is to a large extent driven by the purification systems, pushing LXe across the TPC with flows up to 0.35\,t/h~\cite{Aprile:2022vux}. It is thus very likely that a re-routing of the purification flow, keeping the purified xenon in the TPC to avoid contact with the surfaces of the outer volume, could already lead to a significant $^{222}$Rn reduction.


\section*{Acknowledgements}
This work was supported by the European Research Council (ERC) grant No.~724320 (ULTIMATE). We thank the staff of the workshops at the Institute of Physics in Freiburg for their support.

\end{document}